\documentclass[twocolumn,showpacs,amsmath,amssymb,prl]{revtex4}
\usepackage{amsxtra}
\usepackage{amssymb}
\usepackage{amsmath}
\usepackage{graphicx}

\newcommand {\rhovec}{\ensuremath \hat{\boldsymbol{\rho}}}

\begin{document}
\title{Secure Communication using Gaussian-State Quantum illumination}
\date{\today}
\author{Jeffrey H. Shapiro}\email{jhs@mit.edu}
\affiliation{Research Laboratory of Electronics, Massachusetts Institute of Technology, Cambridge, Massachusetts 02139, USA}
  
\begin{abstract} 
A new paradigm for secure communication, based on quantum illumination, is proposed.  Alice uses spontaneous parametric down-conversion to send Bob a set of signal modes over a pure-loss channel while retaining the set of idler modes with which they are maximally entangled.  Bob imposes a single information bit on the modes he receives from Alice via binary phase-shift keying.  He then adds classical Gaussian noise and sends the noisy modulated modes back to Alice over the same pure-loss channel.  Even though the loss and noise destroy any entanglement between the modes that Alice receives from Bob and the idler modes she has retained, she can decode Bob's bit with an error probability that can be orders of magnitude lower than what is achieved by a passive eavesdropper who receives all the photons that are lost en route from Alice to Bob and from Bob to Alice.   
  
\end{abstract}

\maketitle

The use of quantum key distribution (QKD) to ensure the security of classical information transmission has moved from its theoretical roots \cite{BB84,Ekert,CV} to a major network demonstration \cite{Vienna}.  The objective of QKD is for two geographically separated users---Alice and Bob---to create a shared set of completely random key bits in manner that precludes an eavesdropper (Eve) from having anything more than an inconsequentially small amount of information about the entire set of key bits.  That such a goal is possible arises from a fundamental quantum mechanical principle:  Eve cannot tap the Alice-to-Bob channel without creating a disturbance on that channel.  

In this Letter we will introduce a new paradigm for secure communication using quantum resources.  Although it can be used to generate a secret key, as in existing QKD systems, the enormous disparity between the bit error probabilities of a passive eavesdropper and the intended receiver make this scheme attractive for direct information transmission.  The basis for this new approach is quantum illumination, specifically the Gaussian-state radar system described in \cite{QI}.  There, the entangled signal and idler outputs from spontaneous parametric down-conversion (SPDC) were shown to afford a substantial error probability advantage---over a coherent-state system of the same average transmitted photon number---when the signal beam is used to irradiate a target region containing a bright thermal-noise bath in which a low-reflectivity object might be embedded, and the idler beam is retained at the transmitter for use in an optimal joint measurement with the light returned from the target region.  This performance advantage is surprising, because the loss and noise combine to destroy any entanglement between the return light and the retained idler.  The origin of this advantage is the stronger-than-classical phase-sensitive cross correlation between the signal and idler produced by SPDC.  When the source is operated in the low-brightness regime, this leads to a phase-sensitive cross correlation between the target return and the retained idler that outstrips any such correlation produced by a classical-state transmitter of the same average transmitted photon number \cite{QI}.  Here, we will turn that capability to the task of secure communication between Alice and Bob, despite the presence of a passive eavesdropper Eve.  

The communication system of interest functions as follows.  Alice uses SPDC to produce $M$ signal-idler mode pairs, with annihilation operators $\{\,\hat{a}_{S_m}, \hat{a}_{I_m} : 1 \le m \le M\,\}$, whose joint density operator $\rhovec_{SI}$ is the tensor product of independent, identically distributed (iid) density operators for each mode pair that are zero-mean, jointly Gaussian states with the common Wigner-distribution covariance matrix 
\begin{equation}
{\boldsymbol \Lambda}_{SI} = \frac{1}{4}\left[\begin{array}{cccc}
S  & 0 & C_q & 0 \\ 
0 & S  & 0 & -C_q \\
C_q & 0 & S & 0 \\ 
0 & -C_q & 0 & S
\end{array}\right], \label{quadent}
\end{equation}
where $S \equiv 2N_S + 1$ and $C_q \equiv 2\sqrt{N_S(N_S+1)}$, and $N_S$ is the average photon number of each signal (and idler) mode \cite{footnote1}.  Alice sends her signal modes to Bob, retaining her idler modes for later use.  Alice-to-Bob transmission occurs over a pure-loss channel \cite{PRL}, so that Bob receives modes whose annihilation operators are
\begin{equation}
\hat{a}_{B_m} = \sqrt{\kappa}\,\hat{a}_{S_m} + \sqrt{1-\kappa}\,\hat{e}_{B_m},\mbox{ for $1\le m \le M$,}
\end{equation}
where the environmental modes, $\{\hat{e}_{B_m}\}$, are in their vacuum states \cite{footnote2}.  Bob  first imposes an  identical, binary phase-shift keyed (BPSK) information bit ($k = 0$ or 1) on each $\hat{a}_{B_m}$, yielding $(-1)^k\hat{a}_{B_m}$.  He then adds iid, zero-mean, isotropic, classical Gaussian noise, $n_{B_m}$, of variance $N_B$ to each $(-1)^k\hat{a}_{B_m}$, and transmits the noisy modulated modes, $\hat{a}_{B_m}' \equiv (-1)^k\hat{a}_{B_m} + n_{B_m}$, back to Alice.  After propagation through the same pure-loss channel, Alice receives modes whose annihilation operators are
\begin{equation}
\hat{a}_{R_m} = \sqrt{\kappa}\,\hat{a}_{B_m}' + \sqrt{1-\kappa}\,\hat{e}_{A_m},\mbox{ for $1\le m \le M$},
\end{equation}
where the $\{\hat{e}_{A_m}\}$ are in their vacuum states.  Given Bob's information bit $k$, we have that $\rhovec_{RI}^{(k)}$, the joint state of Alice's $\{\hat{a}_{R_m},\hat{a}_{I_m}\}$ modes, is the tensor product of iid, zero-mean, jointly Gaussian states for each mode pair with the common Wigner-distribution covariance matrix
\begin{eqnarray}
\lefteqn{\hspace*{-.1in}{\boldsymbol \Lambda}^{(k)}_{RI} = } \nonumber \\ 
&&\hspace*{-.2in} \frac{1}{4} \left[\begin{array}{cccc}
A  & 0 & (-1)^kC_a & 0 \\ 
0 & A  & 0 & (-1)^{k+1}C_a \\
(-1)^kC_a & 0 & S & 0 \\ 
0 & (-1)^{k+1}C_a & 0 & S
\end{array}\right]\!\!, \label{quadentrcv}
\end{eqnarray}
where $A \equiv 2\kappa^2 N_S + 2\kappa N_B + 1$ and $C_a \equiv \kappa C_q$.  Alice's task is to decode Bob's bit, which is equally likely to be $k=0$ or $k=1$, with minimum error probability.  

Eve will be assumed to collect \em all\/\rm\ the photons that are lost en route from Alice to Bob and from Bob to Alice \cite{footnote3}, i.e., she has at her disposal the mode pairs $\{\,\hat{c}_{S_m}, \hat{c}_{R_m} : 1 \le m \le M\,\}$, where
\begin{eqnarray}
\hat{c}_{S_m} &=& \sqrt{1-\kappa}\,\hat{a}_{S_m} - \sqrt{\kappa}\,\hat{e}_{B_m},\\
\hat{c}_{R_m} &=& \sqrt{1-\kappa}\,\hat{a}_{B_m}' -\sqrt{\kappa}\,\hat{e}_{A_m}.
\end{eqnarray}
Given Bob's bit value, Eve's joint density operator, $\rhovec_{c_Sc_R}^{(k)}$, is the tensor product of $M$ iid mode-pair density operators that are zero-mean, jointly Gaussian states with the common Wigner-distribution covariance matrix
\begin{equation}
{\boldsymbol \Lambda}^{(k)}_{c_Sc_R} 
= \frac{1}{4}
 \left[\begin{array}{cccc}
D  & 0 & (-1)^kC_e & 0 \\ 
0 & D  & 0 & (-1)^kC_e \\
(-1)^kC_e & 0 & E & 0 \\ 
0 & (-1)^kC_e & 0 & E
\end{array}\right]\!\!, \label{quadenteve}
\end{equation}
where $D \equiv 2(1-\kappa)N_S + 1$, $C_e \equiv 2(1-\kappa)\sqrt{\kappa}\,N_S$, and $E \equiv 2(1-\kappa)\kappa N_S + 2(1-\kappa)N_B + 1$.  Eve too is interested in minimum error-probability decoding of Bob's bit.  

Alice's minimum error probability decision rule is to measure $\rhovec_{SI}^{(1)} - \rhovec_{SI}^{(0)}$, and declare that $k = 1$ was sent if and only if her measurement outcome is non-negative.  Similarly, Eve's minimum error probability decision rule is to measure $\rhovec_{c_Sc_R}^{(1)} -\rhovec_{c_Sc_R}^{(0)}$ and declare that $k=1$ was sent if and only if her measurement outcome is non-negative.  As noted for the quantum illumination radar problem treated in \cite{QI}, the exact error probabilities for these Gaussian-state hypothesis tests are not easy to evaluate.  Thus, as in \cite{QI}, we shall rely on quantum Chernoff bounds \cite{Chernoff}, which are known to be exponentially tight for iid $M$ mode-pair problems, i.e., 
with $\Pr(e) \le e^{-M\max_{0\le s\le 1}{\cal{E}}(s)}/2$,
for ${\cal{E}}(s) \equiv -\ln\!\left({\rm tr}[(\hat{\rho}_m^{(0)})^s(\hat{\rho}_m^{(1)})^{1-s}]\right)$,
giving the Chernoff bound (in terms of the conditional mode-pair density operators $\hat{\rho}_m^{(k)}$) on the exact error probability,
we have
\begin{equation}
\lim_{M\rightarrow \infty}\ln[2\Pr(e)]/M = \max_{0\le s\le 1}{\cal{E}}(s).
\end{equation}
The BPSK symmetry in $\rhovec_{SI}^{(k)}$ and $\rhovec_{c_Sc_R}^{(k)}$ implies that $s=1/2$ optimizes the Chernoff bound exponents for both Alice and Eve.   The following lower bound on the error probability of any receiver \cite{QI} will also be of use:
\begin{equation}
\Pr(e) \ge \frac{1-\sqrt{1-e^{-2M{\cal{E}}(1/2)}}}{2};
\label{lowerbound}
\end{equation}
it is not exponentially tight for the problems at hand.  

Because all our conditional density operators are zero-mean Gaussian states, we can use the results of \cite{Pirandola} to evaluate ${\cal{E}}(1/2)$ for Alice and Eve's receivers.  To do so we need the symplectic diagonalizations of their conditional Wigner-distribution covariance matrices.  The symplectic diagonalization  of a $4\times 4$ dimensional covariance matrix $\boldsymbol \Lambda$ consists of a $4\times 4$ dimensional symplectic matrix $\boldsymbol S$ and a symplectic spectrum $\{\,\nu_n : 1\le n \le 2\,\}$ that satisfy
\begin{eqnarray}
{\boldsymbol S}{\boldsymbol \Omega}{\boldsymbol S}^T &=& {\boldsymbol \Omega} 
\equiv \left[\begin{array}{cccc} 0 &1 & 0 &  0\\ -1 & 0 & 0 & 0 \\ 
0 & 0& 0 & 1 \\ 0 & 0 & -1 & 0 \end{array}\right]   \\[.1in]
{\boldsymbol \Lambda} &=& {\boldsymbol S}\,{\rm diag}(\nu_1,\nu_1,\nu_2,\nu_2)
{\boldsymbol S}^T,
\end{eqnarray} 
where diag($\cdot,\cdot,\cdot,\cdot)$ denotes a diagonal matrix with the given diagonal elements.  

For our quantum-illumination (Alice-to-Bob-to-Alice) communication, the symplectic matrices needed for the diagonalization of ${\boldsymbol \Lambda}^{(k)}_{RI}$ are
\begin{equation}
{\boldsymbol S}^{(k)} = \left[\begin{array}{cc}
{\bf X}_+ & (-1)^k{\bf X}_- \\[.1in] (-1)^k{\bf X}_- & {\bf X}_+\end{array}\right],
\end{equation}
for $k = 0,1$.  Here, 
${\bf X_\pm} \equiv {\rm diag}(x_\pm,\pm x_\pm)$ with 
\begin{equation}
x_\pm \equiv \sqrt{\frac{A+S \pm\sqrt{(A+S)^2 - 4C_a^2}}
{2\sqrt{(A+S)^2 - 4C_a^2}}}.
\end{equation}
 The associated symplectic spectra are identical for $k = 0$ and 1, i.e., for $n=1,2$ we have
\begin{equation}
\nu_n^{(k)} = \left[(-1)^n(S - A) + \sqrt{(A+S)^2 - 4C_a^2}\right]/8.
\end{equation}

For Eve's attempt to listen in, the symplectic matrices needed for the diagonalization of ${\boldsymbol \Lambda}^{(k)}_{c_Sc_R}$ are
\begin{equation}
{\boldsymbol S}^{(k)} = \left[\begin{array}{cc}
{\bf Y} & (-1)^{k+1}{\bf Z} \\[.1in] (-1)^k{\bf Z} & {\bf Y}\end{array}\right],
\end{equation}
for $k = 0,1$.  Here, 
${\bf Y} \equiv {\rm diag}(\cos(\theta),\cos(\theta))$ and ${\bf Z} \equiv {\rm diag}(\sin(\theta), \sin(\theta))$ with 
\begin{equation}
\cos(2\theta) = \frac{D-E}{\sqrt{(D-E)^2 + 4C_e^2}}.
\end{equation}
The associated symplectic spectra are identical for $k=0$ and 1, i.e., for $n=1,2$ we have
\begin{equation}
\nu_n^{(k)} = \left[(D + E) - (-1)^n\sqrt{(D-E)^2 + 4C_e^2}\right]/8.
\end{equation}

The preceding diagonalizations lead to Chernoff bound expressions that are far too long to exhibit here.  In Fig.~1 we compare the Chernoff bounds for Alice and Eve's optimum quantum receivers when $\kappa = 0.1, N_S = 0.004$, and $N_B  = 100$.  Also included in this figure is the error-probability lower bound from (\ref{lowerbound}) on Eve's optimum quantum receiver.  We see that Alice's error probability \em upper\/\rm\ bound---at a given $M$ value---can be orders of magnitude lower than the Eve's error probability \em lower\/\rm\ bound when both use optimum quantum reception.  This occurs despite Eve's getting 9 times more of Alice's transmission than Bob does and 9 times more of Bob's transmission than Alice does.  Note that Alice's performance advantage may be better assessed from comparing her error-probability upper bound with that of Eve's receiver, in that both are exponentially tight Chernoff  bounds.  
\begin{figure}[h]
\begin{center}
\includegraphics[width=2.25in]{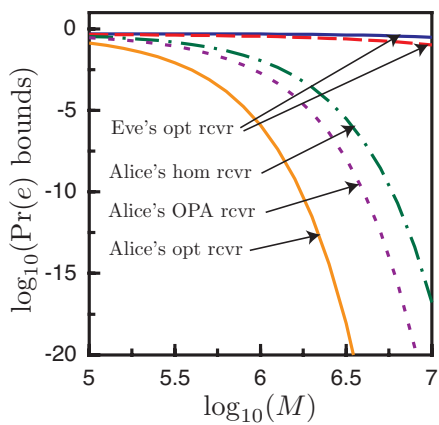}
\end{center}
\vspace*{-.2in}
\caption{(Color online) Error-probability bounds versus $M$, the number of SPDC mode pairs, assuming $N_S = 0.004$, $\kappa=0.1$, and $N_B = 100$.  Solid curves: Chernoff bounds for Alice and Eve's optimum quantum receivers.  Long-dashed curve: error-probability lower bound for Eve's optimum quantum receiver.  Dot-dashed curve:  Chernoff bound for Alice's homodyne receiver.  Short-dashed curve: Bhattacharyya bound for Alice's optical parametric amplifier (OPA) receiver.}
\end{figure}

To show that the advantage afforded by quantum illumination extends well beyond the specific example chosen for Fig.~1, we have used an algebraic computation program to obtain the following approximate forms for the Chernoff bounds on the error probabilities of Alice and Eve's optimum quantum receivers:
\begin{eqnarray}
\Pr(e)_{\rm Alice} &\le& \frac{\exp(-4M\kappa N_S/N_B)}{2} 
\label{AliceOpt}\\  
\Pr(e)_{\rm Eve} & \le& \frac{\exp(-4M\kappa(1-\kappa)N_S^2/N_B)}{2},
\label{EveOpt}
\end{eqnarray}
which apply in the low-brightness, high-noise regime, viz., when $N_S \ll 1$ and $\kappa N_B \gg 1$.  We see that Alice's Chernoff bound error exponent will be orders of magnitude \em higher\/\rm\ than that of Eve in this regime, because
\begin{equation}
{\cal{E}}_{\rm Alice}(1/2)/{\cal{E}}_{\rm Eve}(1/2) = 1/(1-\kappa)N_s \gg 1.
\end{equation}
Thus the advantageous quantum illumination behavior shown in Fig.~1 is typical for this regime.

As yet we have not identified specific implementations for Alice or Eve's optimum quantum receivers.  So, while we will accord Eve the right to an optimum quantum receiver,  let us show that Alice can still enjoy an enormous advantage in error probability when she uses optical homodyne detection \cite{PartIII} to measure $\{\,{\rm Re}(\hat{a}_{R_m}) : 1\le m \le M\,\}$ and $\{\,{\rm Re}(\hat{a}_{I_m}) : 1 \le m \le M\,\}$.  Conditioned on Bob's information bit, these homodyne measurements yield $M$ iid pairs of zero mean, real-valued, jointly Gaussian random variables with common covariance matrix 
\begin{equation}
{\boldsymbol \Lambda}_{\rm hom}^{(k)} = \frac{1}{4}\left[\begin{array}{cc}A & (-1)^kC_a \\ 
(-1)^kC_a & S \end{array}\right]
\end{equation}
Using the classical Chernoff bound \cite{VT}, we then find that 
\begin{equation}
\Pr(e)_{\rm hom} \le \frac{\exp(-M\kappa N_S/N_B)}{2},
\label{homodyne}
\end{equation}
is an exponentially-tight upper bound on the Alice's homodyne-reception error probability in the low-brightness, high-noise regime.   Comparing (\ref{homodyne}) with (\ref{AliceOpt}) we see that 6\,dB of error exponent has been lost by retreating from optimum quantum reception to homodyne detection.  A more effective receiver implementation can be developed from Guha's optical parametric amplifier (OPA) receiver for the quantum-illumination radar \cite{Guha}.  Here Alice uses an OPA to obtain modes given by
\begin{equation}
\hat{a}'_m \equiv \sqrt{G}\,\hat{a}_{I_m} + \sqrt{G-1}\,\hat{a}_{R_m}^\dagger,\mbox{ for $1\le m \le M$,}
\end{equation}
where $G = 1 + N_S/\sqrt{\kappa N_B}$, and then makes a minimum error-probability decision based on the results of the photon-counting measurement $\sum_{m=1}^M\hat{a}'^\dagger_m\hat{a}'_m$.  The Bhattacharyya bound \cite{footnote4} on this receiver's error probability in the $N_S \ll 1$, $\kappa N_B \gg 1$ regime turns out to be
\begin{equation}
\Pr(e)_{\rm OPA} \le \frac{\exp(-2M\kappa N_S/N_B)}{2},
\label{OPA}
\end{equation}
which is only 3\,dB inferior, in error exponent, to Alice's optimum quantum receiver.  We have included the numerically-evaluated error probability bounds for Alice's homodyne (Chernoff bound) and OPA (Bhattacharyya bound) receivers in Fig.~1, for the case $\kappa = 0.1$, $N_S = 0.004$, and $N_B=100$.    

We have demonstrated that quantum illumination offers a new approach to secure communication in the lossy ($\kappa \ll 1$), noisy ($\kappa N_B \gg 1$), low-brightness ($N_S \ll 1$) regime.  In Fig.~2  we show that high noise and low brightness are essential to this communication scheme by comparing the Chernoff bounds for Alice and Eve's optimum quantum receivers when $\kappa = 0.1, N_S = 0.004$, and $N_B  = 0$, and when $\kappa = 0.1, N_S = 10$, and $N_B = 100$.  In the former situation, quantum-illumination reception performs \em worse\/\rm\ than eavesdropping, because of  Eve's collecting the lion's share of the photons sent by Alice and by Bob.  In the latter case, quantum-illumination reception is almost equivalent to eavesdropping, because the phase-sensitive cross-correlation from high-brightness SPDC is only slightly stronger than the classical limit.    
\begin{figure}[h]
\begin{center}
\includegraphics[width=2.25in]{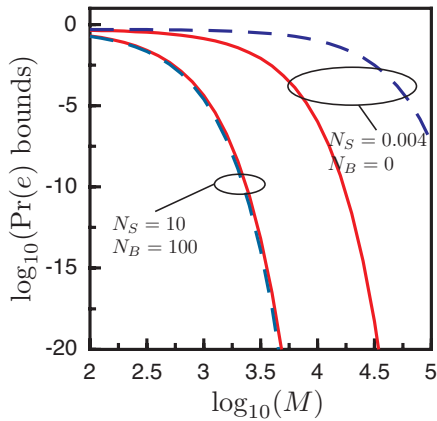}
\end{center}
\vspace*{-.2in}
\caption{(Color online) Chernoff bounds versus $M$, the number of SPDC mode pairs, for the no-noise and high-brightness regimes.   Optimum quantum reception and $\kappa = 0.1$ is assumed for Alice (dashed curves) and Eve (solid curves).}
\end{figure}

Some final points are worth noting.  BPSK communication is intrinsically phase sensitive, so Alice's receiver will require phase coherence that must be established through a tracking system.  Quantum-illumination secure communication will require authentication, lest Eve insert herself between Alice and Bob---in a man-in-the-middle attack---pretending to be Bob to Alice and Alice to Bob.  Authentication might be carried out over a classical optical communication link, as is done in QKD systems, and could be augmented by checks on the physical integrity of the Alice-to-Bob connection, e.g., using optical time-domain reflectometry on a fiber link.  Finally, there is the path-length versus bit-rate tradeoff.  Operation must occur in the low-brightness regime.  So, as channel loss increases ($\kappa$ decreases), Alice must increase her mode-pair number $M$ at constant $N_S$ to maintain a sufficiently low error probability \em and\/\rm\ communication security.  If she uses $T$-sec-long time intervals for each bit, i.e., a bit rate of $R = 1/T$, with an SPDC source of $W$\,Hz phase-matching bandwidth, then $M = WT$ \cite{WJ} implies that her bit rate will go down as loss increases and error probability is held constant.  For the case shown in Fig.~1, we note that a 1\,THz phase-matching bandwidth and 2\,$\mu$s bit duration will yield highly-secure 500\,kbit/s communication---$\Pr(e)_{\rm OPA} \le 7.15 \times 10^{-6}$ and $0.285 \le \Pr(e)_{\rm Eve} \le 0.451$---with $M = WT = 2 \times 10^6$ when Alice and Bob are linked by 50\,km of  0.2\,dB/km loss fiber, assuming that the rest of their equipment is ideal.  

In conclusion, quantum illumination can provide communication that is secure against passive eavesdropping in an entanglement-destroying environment.  Additional steps will be needed to defeat active attacks, in which Eve uses her own light to probe Bob's phase modulator.

This research was supported by the Office of Naval Research Basic Research Challenge Program, the W. M. Keck Foundation Center for Extreme Quantum Information Theory and the DARPA Quantum Sensors Program.


\begin{thebibliography}{2}
\bibitem{footnote1}This state has maximum quadrature entanglement for its average photon number.  Its closest classical-state counterpart is the zero-mean, jointly Gaussian state whose Wigner-distribution covariance is given by (\ref{quadent}) with $C_q$ replaced by $C_c \equiv 2N_S$.  For low-brightness operation, wherein $N_S \ll 1$ prevails, we have $C_q \gg C_c$.
\bibitem{footnote2}At near-infrared through ultraviolet wavelengths, background light will be sufficiently weak in both fiber and free-space channels that it can be ignored.  For example, a typical daytime spectral radiance value of 10\,W/m$^2$\,sr\,$\mu$m at 1.55\,$\mu$m wavelength \cite{Bordogna} leads to $\langle \hat{e}_{B_m}^\dagger \hat{e}_{B_m}\rangle \sim 10^{-6}$ for a line-of-sight terrestrial link;  nighttime values are several orders of magnitude lower.  Our noiseless channel model will suffice so long as $N_S \gg 10^{-6}$.
\bibitem{footnote4}The Bhattacharyya bound is the Chernoff bound with $s=1/2$ used even when it is not the optimum choice.
\end{thebibliography}
\end{document}